\documentclass[10pt,preprint]{aastex}
\shorttitle{Structure of the Homunculus}
\shortauthors{Smith}
\begin{document}

\title{THE STRUCTURE OF THE HOMUNCULUS: I. SHAPE AND LATITUDE
  DEPENDENCE FROM H$_2$ AND [Fe~{\sc ii}] VELOCITY MAPS OF ETA
  CARINAE\altaffilmark{1}}

\author{Nathan Smith\altaffilmark{2}}
\affil{Center for Astrophysics and Space Astronomy, University of
Colorado, 389 UCB, Boulder, CO 80309}

\altaffiltext{1}{Based on observations obtained at the Gemini
Observatory, which is operated by AURA, under a cooperative agreement
with the NSF on behalf of the Gemini partnership: the National Science
Foundation (US), the Particle Physics and Astronomy Research Council
(UK), the National Research Council (Canada), CONICYT (Chile), the
Australian Research Council (Australia), CNPq (Brazil), and CONICET
(Argentina).}

\altaffiltext{2}{Hubble Fellow; nathans@casa.colorado.edu}

\begin{abstract}

High resolution long-slit spectra obtained with the Phoenix
spectrograph on Gemini South provide our most accurate probe of the
three dimensional structure of the Homunculus Nebula around
$\eta$~Carinae.  The new near-infrared spectra dramatically confirm
the double-shell structure inferred previously from thermal dust
emission, resolving the nebula into a very thin outer shell seen in
H$_2$ $\lambda$21218, and a warmer, thicker inner layer seen in
[Fe~{\sc ii}] $\lambda$16435.  The remarkably thin and uniform H$_2$
skin has $\Delta$R/R of only a few per cent at the poles, hinting that
the most important mass loss during the 19th century eruption may have
had a very short duration of $\la$5 yr.  H$_2$ emission traces the
majority of the more than 10 M$_{\odot}$ of material in the nebula,
and has an average density of order $n_H\ga$10$^{6.5}$ cm$^{-3}$.
This emission, in turn, yields our first definitive picture of the
exact shape of the nebula, plus a distance of 2350$\pm$50 pc and an
inclination angle of $\sim$41\arcdeg\ (the polar axis is tilted
49\arcdeg\ from the plane of the sky).  The distribution of the H$_2$
emission provides the first measure of the latitude dependence of the
speed, mass loss, and kinetic energy associated with $\eta$~Car's 19th
century explosion.  Almost 75\% of the total mass and more than 90\%
of the kinetic energy in the ejecta were released at high latitudes
between 45\arcdeg\ and the polar axis.  This rules out a model for the
bipolar shape wherein an otherwise spherical explosion was pinched at
the waist by a circumstellar torus.  Also, the ejecta could not have
been deflected toward polar trajectories by a companion star, since
the kinetic energy of the polar ejecta is greater than the binding
energy of the putative binary system.  Instead, most of the mass
appears to have been directed poleward by the explosion itself --- or
the star failed to launch material from low latitudes, which would
have important consequences for the angular momentum evolution of the
star.  In any case, comparing H$_2$ and [Fe~{\sc ii}] emission
resolves some puzzles about structure noted in previous studies.
H$_2$ emission also provides our first reliable picture of the
critical innermost waist of the Homunculus, yielding clues to the
observed morphology of the core and the more extended equatorial
debris.

\end{abstract}

\keywords{circumstellar matter --- ISM: individual (Homunculus Nebula)
  --- stars: individual ($\eta$ Carinae) --- stars: mass loss ---
  stars: winds, outflows}

\section{INTRODUCTION}

The Homunculus Nebula around $\eta$ Carinae is composed of a striking
pair of lumpy bipolar lobes divided by a ragged equatorial skirt
(Morse et al.\ 1998; Duschl et al.\ 1995; Smith et al.\ 2000; Smith \&
Gehrz 2000).  It is primarily a reflection nebula at UV, visual, and
near-IR wavelengths, since it is highly polarized (Thackeray 1956b,
1961; Visvanathan 1967; Meaburn et al.\ 1993; Schulte-Ladbeck et al.\
1999), and because stellar wind emission lines are seen reflected by
the nebula (Allen \& Hillier 1991; Davidson et al.\ 1995; Smith et
al.\ 2003a; Stahl et al.\ 2005).  While spectacular in appearance,
scattered-light images do not give a clear picture of the three
dimensional (3D) shape, tilt angle, or detailed structure of the
nebula because the optical depth is near unity and the inclination
angle hides parts of the nebula.  Additionally, the underlying
structure giving rise to the clumpy, mottled appearance of the polar
lobes in {\it Hubble Space Telescope} ({\it HST}) and infrared (IR)
images defies easy explanation.  Different datasets imply different
mass distributions in dark lanes, holes, and bright cells or clumps
(e.g., Morse et al.\ 1998; Smith et al.\ 1998), evoking entertaining
comparisons to structures seen in convection cells and vegetables.

Because of its prominence and a wealth of high-quality observational
data, the Homunculus is an important prototype of bipolar nebulae,
being the focus of numerous theoretical studies (Frank et al.\ 1995,
1998; Owocki \& Gayley 1997; Dwarkadas \& Balick 1998; Langer et al.\
1999; Dwarkadas \& Owocki 2002; Owocki 2003; Soker 2001, 2004; Matt \&
Balick 2004; Gonzalez et al.\ 2004a, 2004b).  Yet, the formation of
its clumpy bipolar lobes remains an enduring mystery.  Even more
perplexing is its ragged equatorial disk, which makes the Homunculus
unique among bipolar nebulae (Davidson et al.\ 1997, 2001; Zethson et
al.\ 1999; Smith \& Gehrz 1998; Smith et al.\ 1998, 2002).  Detailed
study of this nebula provides one of our most valuable ways to learn
about the Great Eruption of $\eta$ Carinae in the mid-19th century,
since proper motions indicate that the nebula was ejected in that
event (Gaviola 1950; Ringuelet 1958; Currie et al.\ 1996; Smith \&
Gehrz 1998; Morse et al.\ 2001).  An important observational goal is
to measure the exact shape and mass distribution in the nebula,
because it bears the direct imprint of the physics of the Great
Eruption as a function of latitude.  This, in turn, can constrain the
roles of rotation, binarity, and the hydrodynamics of interacting
winds in the formation of bipolar structure.

Thermal-IR observations have yielded important clues about the
structure of the Homunculus, because the nebula is optically thin at
these wavelengths (they have also provided our best estimates of its
total mass of more than 10 M$_{\odot}$ and IR luminosity of
4.3$\times$10$^6$ L$_{\odot}$; Smith et al.\ 2003b).  Mid-IR maps by
Hackwell, Gehrz, \& Grasdalen (1986) first revealed that the bipolar
nebula consisted of a pair of hollow, limb-brightened osculating
spheroids.  This basic structure was later supported by a much higher
resolution monochromatic 12 $\micron$ image (Smith et al.\ 1995),
multiwavelength mid-IR images (Smith et al.\ 1998; Polomski et al.\
1999), mid-IR imaging polarimetry (Aitken et al.\ 1995), and
spectroscopy (Hillier \& Allen 1992; Meaburn et al.\ 1993).  While the
Homunculus had roughly the same size in thermal emission from 8 to 13
$\micron$, Smith et al.\ (1998) noted that it was considerably larger
in scattered light images at $\lambda < 2 \ \micron$.  This mystery
was resolved with higher resolution thermal-IR imaging using larger
telescopes.  Smith et al.\ (2003b) used images from the Magellan
Observatory to demonstrate that the nebula was also larger at 18--25
$\micron$ than at 8--13 $\micron$.  In that paper, we proposed a {\it
double-shell structure} for the polar lobes, wherein a warm ($\sim$200
K) inner shell dominates the 8--13 $\micron$ images, and a cooler
($\sim$140 K) outer skin dominates at longer wavelengths.  The
double-shell structure will be an important recurring theme in the
present investigation.  This model also indicated that most of the
$>$10 M$_{\odot}$ in the Homunculus was associated with the cool outer
skin of the polar lobes, settling an earlier debate about the origin
of the far-IR emission from $\eta$ Car (Davidson \& Smith 2000; Morris
et al.\ 1999).  Thermal-IR maps show complex dust structure in the
homunculosity core as well (Smith et al.\ 1995, 2002; Chesneau et al.\
2005), which will be discussed in detail at the end of this paper.

Since the Homunculus is essentially a ballistic expanding Hubble flow,
measured Doppler shifts in the ejecta provide another viable method to
trace the 3D shape.  Several authors have utilized this technique with
spectroscopy of visual-wavelength emission lines (Thackeray 1951,
1956a, 1956b, 1961; Meaburn et al.\ 1987, 1993; Hillier \& Allen 1992;
Allen \& Hillier 1991, 1993; Currie et al.\ 1996; Meaburn 1999), as
well as imaging polarimetry (Schulte-Ladbeck et al.\ 1999).  However,
none of these various studies gave a definitive measure of the shape,
largely due to the low spatial resolution of ground-based
spectroscopy, plus the very bright scattered continuum light and high
optical depth of the dusty nebula at visual wavelengths (see Hillier
1997 for a review of proposed model shapes).  This situation improved
significantly when high spatial resolution optical spectra from the
Space Telescope Imaging Spectrograph (STIS) became available.
Davidson et al.\ (2001) studied the kinematic structure of narrow
[Ni~{\sc ii}] and [Fe~{\sc ii}] emission lines in STIS spectra to give
the first clear picture of the shape of the polar lobes and their
inclination angle, plus the first genuine detection of the thin
equatorial disk.  However, our major result in that study was still
rather unsatisfying: our model shape that best fit the kinematic data
(``Model 1'' of Davidson et al.\ 2001) yielded a projected image of
the polar lobes that was {\it far too small and skinny} to explain
images of the Homunculus.  We shall see below that the solution to
this mystery is that the [Ni~{\sc ii}] and [Fe~{\sc ii}] lines studied
at visual wavelengths actually trace a smaller inner shell of the
polar lobes, and not the dominant outer layer of the Homunculus seen
in images.

The discovery of molecular gas in the Homunculus (Smith 2002a; Smith
\& Davidson 2001) provided a powerful new tool to study its structure
via near-IR emission from H$_2$ $v$=1--0 S(1) $\lambda$21218.  This
emission line is bright and narrow, is only seen in intrinsic emission
from the polar lobes (no confusing reflected emission), and is at a
wavelength of $\sim$2~$\micron$ where we can begin to see through to
the back of the Homunculus.  Hence, spectra of H$_2$ revealed emission
from the far side of the Homunculus lobes for the first time (Smith
2002a).  Smith (2002a) showed that the H$_2$ emission most likely
originates in an outer shell, separate from the inner [Fe~{\sc ii}]
shell.  Thus, H$_2$ probably traces gas associated with the same outer
cool dust shell seen in thermal-IR images (Smith et al.\ 2003b).  It
is therefore our most powerful kinematic diagnostic of the majority of
the mass in the Homunculus, and traces the component of the nebula
that dominates images.  However, earlier [Fe~{\sc ii}] and H$_2$
spectra had inadequate spectral resolution to clearly discern the
relative thickness of the two shells or their detailed structure.  The
present study aims to rectify this situation using long-slit spectra
of H$_2$ and [Fe~{\sc ii}] obtained with much higher dispersion.

\section{OBSERVATIONS}

High-resolution ($R\simeq$60,000; $\sim$5 km s$^{-1}$) near-IR spectra
of $\eta$ Car were obtained on 2005 March 1 and 2 using the Phoenix
spectrograph (Hinkle et al.\ 2003) on the Gemini South telescope.
Phoenix has a 1024$\times$256 InSb detector with a pixel scale of
0$\farcs$085$\times$1.4 km s$^{-1}$ at a wavelength of
$\sim$1.6~$\mu$m, and 1.26 km s$^{-1}$ at $\sim$2~$\micron$.  Sky
conditions were photometric, and the seeing was typically
0$\farcs$3--0$\farcs$5 (better at the longer wavelength).  Removal of
airglow lines was accomplished by subtracting an observation of an
off-source position 35$\arcsec$ southeast of the star.  The
0$\farcs$25-wide long-slit aperture was oriented at P.A.=310$\arcdeg$
along the polar axis of the Homunculus.  To sample the kinematics
across the nebula, the slit was positioned on the bright central star,
plus offsets of 1$\arcsec$ and 2$\arcsec$ in either direction
perpendicular to the slit axis as shown in Figure 1.  The observations
were very similar to those described in earlier papers presenting
Phoenix data (Smith 2004, 2005).

HR~3685 and HR~4621 were observed with Phoenix on the same nights with
the same grating settings in order to correct for telluric absorption
and for flux calibration.  Telluric lines were used for wavelength
calibration, using the telluric spectrum available from NOAO.
Velocities were calculated adopting a vacuum rest wavelength of
16439.981 \AA \ for the [Fe~{\sc ii}] ($a^4F-a^4D$) line, and
21218.356 \AA \ for H$_2$ $v$=1--0 S(1).  These velocities were
corrected to a heliocentric reference frame; heliocentric velocities
will be quoted here.  Uncertainty in the resulting velocities is
$\pm$1 km s$^{-1}$, dominated by scatter in the dispersion solution
for telluric lines.

At all five slit placements, long-slit spectra were obtained for the
two emission lines: [Fe~{\sc ii}] $\lambda$16435 and H$_2$
$\lambda$21218.  These were suspected to be the most unambiguous
tracers of the structure in the polar lobes of the Homunculus at
near-IR wavelengths that can penetrate the dusty nebula (Smith 2002a).
The Phoenix slit is shorter than the $\sim$18\arcsec\ extent of the
Homunculus, so each offset position in Figure 1 actually consisted of
exposures at three steps along the slit that were shifted and combined
later to yield a greater effective slit length.  At each pointing, two
or three exposures, each consisting of 2$\times$60-second integrations
(to avoid severe saturation) were obtained.  Depending on the overlap
at various positions along the slit, the total effective integration
time after adding all exposures for a given slit position varied from
240 to 1080 seconds.  Earlier experience with Phoenix (Smith 2004,
2005) revealed that the maximal extent of the blue and redshifted
H$_2$ velocities in the Homunculus reached slightly beyond the usable
bandwith of a single Phoenix exposure.  Therefore, for H$_2$ emission,
observations were obtained at two different grating settings separated
by about 400 km s$^{-1}$; one optimized for including the structure of
the blushifted polar lobe and one for the redshifted lobe.  The
exposures at the redshifted setting were then scaled to have exactly
the same dispersion per pixel, and combined with the blueshifted data.
(Since the extreme redshifted velocities are only seen in the
northwest polar lobe, the southeast pointings along the slit were
ignored.  This explains the data gap in the lower right corner of each
panel on the right side of Fig.\ 2.)

Figure 2 shows the resulting long-slit data for [Fe~{\sc ii}] and
H$_2$, respectively, at the five different slit positions, where a
linear fit to the bright reflected continuum light in the Homunculus
has been subtracted out to enhance the contrast of the line emission
(see Smith 2002a, 2004, 2005 for more details).  The vertical dashed
lines in Figure 2 mark the systemic velocity of $\eta$ Car at --8.1 km
s$^{-1}$, measured from earlier Phoenix spectra of H$_2$ in the
Homunculus (Smith 2004).

\section{THE DOUBLE-SHELL MORPHOLOGY}

Three distinct kinematic components can be seen in Figures 2 and 3: 1)
A thin outer shell of H$_2$ at $\pm$600 km s$^{-1}$ and $\pm$8\arcsec,
2) a thicker inner skin of [Fe~{\sc ii}] at $\pm$500 km s$^{-1}$ and
$\pm$8\arcsec, and 3) [Fe~{\sc ii}] emission at $\pm$250 km s$^{-1}$
and $\pm$2\arcsec\ from the smaller bipolar nebula known as the
``Little Homunculus'', which was the subject of a previous paper
(Smith 2005).\footnote{The Little Homunculus appears ``stretched''
horizontally compared to the larger Homunculus in [Fe~{\sc ii}]
emission in Figs.\ 2 and 3.  This is because it is younger and thus
follows a different Hubble law than the larger Homunculus (Smith 2005;
Ishibashi et al.\ 2003).}

It is clear from Figure 3 that H$_2$ and [Fe~{\sc ii}] emission in the
polar lobes do not share the same spatial distribution.  In fact, they
are strongly anticorrelated.  The observed kinematic structure gives a
dramatic and independent confirmation of the double-shell structure
proposed intially from thermal-IR images (Smith et al.\ 2003b),
consisting of a warm inner dust shell, plus a cooler and denser outer
shell.  Here, the [Fe~{\sc ii}] emission clearly traces the warmer
inner shell, while the H$_2$ emission corresponds to the more massive
thin outer skin.

\subsection{The Inner [Fe~{\sc ii}] Shell}

[Fe~{\sc ii}] emission in Figures 2 and 3 coincides with the warmer
inner dust shell of the polar lobes, and therefore traces only about
10\% of the total mass in the Homunculus (Smith et al.\ 2003b).  For
this reason, the main focus of the present paper will be on the H$_2$
emission.  However, a few comments are in order, as the kinematic
structure of the inner [Fe~{\sc ii}] shell is closely related to the
more massive cool H$_2$ skin.

At all latitudes, [Fe~{\sc ii}] emission from the lobes has a smaller
radius than the H$_2$ emission and clearly represents a separate
component of the nebula.  The infrared [Fe~{\sc ii}] emission shown
here is emitted by the same gas that gives rise to optical [Ni~{\sc
ii}] and [Fe~{\sc ii}] lines.  This explains why Davidson et al.\
(2001) found that their shape for the Homunculus derived from spectra
was too small and skinny compared to images.

The inner [Fe~{\sc ii}] layer and the double-shell structure may also
hold clues for understanding polarization studies of the Homunculus.
While dust in the outer H$_2$ skin traces the main scattering layer
seen in visual images, dust associated with the inner [Fe~{\sc ii}]
zone may contribute significantly to the observed polarization.  If
the outer layer contains predominantly large grains that behave like
black-bodies at thermal-IR wavelengths, and the inner [Fe~{\sc ii}]
layer contains smaller grains with higher temperatures (as observed)
that are seen through holes in the clumpy H$_2$ skin, it may go a long
way toward explaining the apparent contradictions in grain properties
inferred independently from thermal-IR emission and visual to near-IR
polarization (Walsh \& Ageorges 2000; Schulte-Ladbeck et al.\ 1999;
Smith et al.\ 2003b).

The [Fe~{\sc ii}] layer is thicker and more irregular than the H$_2$
shell, partially filling the interior of the lobes, especially evident
in Figure 3.  Its origin and relationship to the more massive H$_2$
shell are not clear.  On one hand, the observed double-shell structure
resembles a classic two-shock structure of a wind plowing into an
ambient medium.  In this interpretation, the [Fe~{\sc ii}] shell might
represent emission from cooling stellar wind material that has passed
through a reverse shock.  The actual shock velocity could be quite low
(less than 100 km s$^{-1}$), since the Homunculus expansion speed at
high latitudes is not much different than the post-eruption stellar
wind speed (Smith et al.\ 2003a).  On the other hand, the [Fe~{\sc
ii}] (and warm dust) layer may simply be a photodissociation region,
where the radiation field from the central star is still too strong to
allow H$_2$ to exist.  Dust in this region attenuates UV light,
eventually allowing H$_2$ to survive at larger radii (see, e.g.,
Ferland et al.\ 2005).  An almost identical double-shell structure of
H$_2$ and [Fe~{\sc ii}] also exists in the famous planetary nebula
M~2-9 (Hora \& Latter 1994; Smith, Balick, \& Gehrz 2005).

At all five slit positions, the [Fe~{\sc ii}] emission from the Little
Homunculus is much brighter in the blueshifted (SE) lobe, while the
[Fe~{\sc ii}] from the larger Homunculus is always brighter in the
redshifted (NW) lobe.  Previously, this had been interpreted as
attenuation of light from the NW lobe of the Little Homunculus as it
passed through dust in the equator (Smith 2005), while the brightness
asymmetry in the larger lobes was caused by the thick polar caps and
thinner side walls of the nebula (Davidson et al.\ 2001; Hillier \&
Allen 1992).  Together, these two explanations have a hint of
inconsistency.  A more plausible conjecture might be that the
redshifted Little Homunculus lobe has a lower density and is therefore
optically thinner in the Balmer continuum, allowing more UV to leak
through to illuminate the larger polar lobe.  A higher density in the
SE polar lobe of the Little Homunculus would explain why it is
brighter in the near-IR, and may help explain the variable extended
radio continuum and predominantly blueshifted radio recombination
lines (Duncan \& White 2003; Duncan et al.\ 1995).  In any case, the
spatial extent of the Little Homunculus (Smith 2005) matches the
extent of variable radio continuum emission (Duncan \& White 2003), so
it would appear that the radio outbursts of $\eta$ Car during the 5.5
yr cycle are due the Little Homunculus absorbing all the Lyman
continuum radiation that escapes from a variable central source.

Some faint [Fe~{\sc ii}] emission from the equatorial skirt is present
as well in panels b and c of Figures 2 and 3, as had been seen
previously in [Fe~{\sc ii}], [Ni~{\sc ii}], [Ca~{\sc ii}], and He~{\sc
i} $\lambda$10830 (Zethson et al.\ 1999; Davidson et al.\ 2001; Smith
2002a; Hartman et al.\ 2004).  The reader should be cautioned that the
curved [Fe~{\sc ii}] emission feature extending from the center to the
lower right in each panel of Figure 2 is not intrinsic emission from
gas in the Homunculus; it is a narrow emission line from the Weigelt
knots near the star that is reflected and redshifted by expanding dust
in the SE polar lobe (see Smith 2002a; Davidson et al.\ 2001).

\subsection{The Thin Outer H$_2$ Shell}

While bright [Fe~{\sc ii}] emission is seen in all nebulae around
luminous blue variables (Smith 2002b), $\eta$ Car is the only one
where H$_2$ emission has been detected.  This unique emission feature
in the Homunculus turns out to be extremely valuable in deducing the
shape and other physical properties of the Homunculus, as detailed in
the rest of this paper.  This is because the H$_2$ emission traces the
cool and dense outer skin of the Homunculus, which thermal-IR
observations indicate contains the majority of the nebula's mass and
kinetic energy (Smith et al.\ 2003b).

Seen at the high spectral resolution provided by the Phoenix
spectrograph on Gemini South, the outer H$_2$ skin is remarkably thin
and uniform.  The thickness of the H$_2$ shell ($\Delta$R) is
typically 300--600 AU (0.5--1$\times$10$^{16}$ cm), or roughly 2--3\%
\ of the polar radius (R).  {\it This gives us an important clue to
the short duration of the most important mass-loss during the Great
Eruption}.  In previous studies, I have commented that the thickness
of the lobes compared to their largest radius ($\Delta$R/R) is about
1\arcsec\ / 8\arcsec, which is the same as the duration of the bright
phase of the Great Eruption compared to the time elapsed since then:
$\Delta$t/t = 20yr/160yr (Smith et al.\ 1998, 2003b; Smith 2002a).
However, now we see that this estimate corresponds to both the inner
and outer portions of the double shell structure, with only a small
fraction of the mass ($\la$10\%) found in the thicker, warmer inner
shell.  The very thin outer shell, where 90\% of the mass resides,
would seem to imply that $\Delta$t/t is only a few per cent, so that
the duration of the dominant mass ejection phase was $\la$5 yr.  This
is supported by the small age dispersion derived from high-precision
proper motion measurements with {\it HST} (e.g., Morse et al.\ 2001).
This, in turn, would imply a {\it huge} instantaneous mass-loss rate
during the peak of the eruption of about 2--5 M$_{\odot}$ yr$^{-1}$.
This may be higher than the limit for a super-Eddington
continuum-driven wind (Owocki et al.\ 2004; Shaviv 2000).  Such a
short duration for the eruption and the corresponding high ratio of
mechanical to radiative energy may point instead to a mechanism more
like a hydrodynamic explosion originating deep within the star (e.g.,
Arnett et al.\ 2005; Young 2005).  However, the very thin H$_2$ skin
also requires a very small velocity dispersion at each latitude during
the original ejection, which is not a property that one usually
associates with an explosion.  A caveat, of course, is that even
though the Homunculus is seen as a Hubble flow today, we cannot be
certain that $\Delta$R/R exactly traces $\Delta$t/t, since thermal
collapse of the shell or ``pile-up'' in the ejecta may have occurred
during or shortly after the eruption.\footnote{It is unlikely,
however, that the thin shell was swept up by the post-eruption stellar
wind long after the eruption, since the characteristic Rayleigh-Taylor
like structures that we might expect in that scenario are absent in
the H$_2$ shell.}  No observations currently exist with which one can
reliably rule out either possibility, so both wind and explosion
models are worth exploring theoretically.

In any case, the outer H$_2$ skin is remarkably thin at the present
epoch --- thinner than had been recognized in previous near-IR
spectroscopy or thermal-IR imaging (Smith 2002a; Smith et al.\ 2003b).
This narrow outermost layer gives rise to a sharp absorption feature
at --513 km s$^{-1}$ where it crosses our line of sight to the central
star, seen in H$_2$ as well as several low-ionization metals like
Ti~{\sc ii} and Fe~{\sc i} in UV spectra of $\eta$~Car (Gull et al.\
2005; Nielsen et al.\ 2005).  Instead of tracing just one sightline,
long-slit near-IR spectra have the advantage of showing emission
patterns from the full Homunculus.  Comparing the UV absorption
profiles of Gull et al.\ (2005) to the spectra in Figure 2, it is
clear that the --513 km s$^{-1}$ component is from the outer H$_2$
skin, the --146 km s$^{-1}$ component arises in the ``Little
Homunculus'' (as noted already by Smith 2002a and Smith 2005), and the
several intermediate velocity components seen in low ionization metals
correspond to the thick lumpy [Fe~{\sc ii}] emitting layer seen in
the near-IR (note that the --513 and --146 km s$^{-1}$ absorption
components are indentified by small white dots on the spectrum with
the slit passing through the star in Figs.\ 2 and 3).

In the discussion in following sections, it will ease matters if it is
fair to assume that the H$_2$ layer can be approximated as a thin skin
of uniform density and constant thickness. The H$_2$ skin obviously
has some clumping and some minor brightness variations, but overall
the shell has a remarkably constant thickness and brightness --
certainly within $\pm$50\%.  One apparent exception to the large-scale
symmetry is in the portions of the walls of the polar lobes that run
tangent to our line of sight, projected near the position of the star.
These portions appear much fainter in Figure 2 than the corresponding
parts of the lobes at the same latitudes (0--40\arcdeg) which are
nearly in the plane of the sky and are seen perpendicular to the
dispersion direction.  These differences give the impression of strong
point symmetry in the brightness distribution at low latitudes in the
Homunculus.  However, it is not certain that this effect is real.  The
problem is that the fainter portions are narrow features stretching
along the dispersion direction, and coincide spatially with very
bright reflected continuum emission that has been subtracted.  It is
quite possible that some H$_2$ emission was subtracted-out when
attempting to fit the continuum at these positions.  Also, these
regions may suffer considerable internal extinction as we look through
the limb of the outer dust shell.  On the other hand, if these
differences mark a real deficit of material, then the possible
departures from a uniform-density model discussed here are in the
sense that there will be additional missing material at low latitudes
in the side walls of the polar lobes, as compared to the assumption of
a uniform skin.  This would only reinforce the general conclusion that
most of the mass is at high latitudes (see \S 5).

\section{DEFINITIVE SHAPE OF THE POLAR LOBES}

Since near-IR H$_2$ emission traces the outer dust shell with most of
the mass in the Homunculus, and the primary dust scattering layer that
dominates continuum images of the reflection nebula (e.g., Davidson \&
Ruiz 1975), it is our most valuable diagnostic of the true shape of
the Homunculus.\footnote{ In hindsight, Ca~{\sc ii} H and K absorption
is also a useful diagnostic, as it seems to be located at slightly
larger radii than the [Fe~{\sc ii}] shell (Davidson et al.\ 2001;
Smith 2002a).  However, it is seen in absorption, tracing only the
near side of each lobe.}  The slit position passing through the
central star is best for this purpose, as it is oriented along the
symmetry axis of the nebula at P.A.$\simeq$310\arcdeg.

The age of the Homunculus is $\sim$160 yr based on proper motion
measurements and on the most dramatic change in brightness during the
Great Eruption (Currie et al.\ 1996; Smith \& Gehrz 1998; Morse et
al.\ 2001; Frew 2004), and the entire outflow is essentially a Hubble
flow.  Therefore, the observed Doppler shift in km s$^{-1}$ ($V_R$)
translates directly to distance from the plane of the sky as
$D_R$=33.5$\times \ V_R$, measured in AU.  Similarly, for a given
distance from Earth, the pixel scale in arcseconds can be converted to
a physical distance in the plane of the sky in order to create a model
shape with the correct physical proportion.  Assuming axial symmetry
then yields an independent estimate of the distance to $\eta$ Car, and
the orientation of the homunculosity (see also Davidson et al.\ 2001).

\subsection{A Symmetric Model Shape}

Figure 4 shows the measured H$_2$ emission shape for a cross section
through the symmetry axis of the Homunculus (with the slit centered on
the star).  The relative scale of the spatial and velocity axes was
adjusted to yield a model shape with the greatest degree of axial
symmetry, and then rotated so that the polar axis is vertical in
Figure 4 (Figs.\ 3f--h were created the same way).  The distance and
inclination angle adopted to make Figure 4 are $D$=2350$\pm$50 pc and
$i$=41$\fdg$0$\pm$0$\fdg$5, respectively.  The inclination angle $i$
given here corresponds to the tilt of the polar axis out of the line
of sight.  It is the same quantity as the inclination defined for
binary system orbits, so that the polar axis is tilted from the plane
of the sky by 49\arcdeg.  The uncertainty is dominated by real
deviations from axial symmetry in the Homunculus, which are of order
2--3\%\ of the radius.  This distance estimate is larger than previous
estimates using the same technique (Smith 2002a; Davidson et al.\
2001, Allen \& Hillier 1993), but the various studies agree to within
the uncertainty.  The distance derived here, however, is the first to
use high-resolution data for the outer H$_2$ shell that is actually
seen in images, instead of the interior [Fe~{\sc ii}] and [Ni~{\sc
ii}] emitting layer.

The best fitting idealized symmetric model for the shape of the polar
lobes is plotted over the data in Figure 4.  This curve provides a
direct measure of the size (radial separation from the star in AU) of
the Homunculus as a function of latitude, which is listed in Table 1.
This is a definitive measure of the true 3D shape of the nebula; it
supersedes previous efforts largely because Doppler shifts of the
narrow molecular hydrogen line constitute such a powerful and unique
tracer of the outer massive shell of the Homunculus.  Earlier studies
using optical [Fe~{\sc ii}] and [Ni~{\sc ii}] were not measuring the
outer shell at all, but instead measured the inner shell, which has an
entirely different shape.  The word ``definitive'' in this context is
not hyperbole, because one can't do much better with observational
data --- uncertanties in the derived model shape are dominated by the
inherent corrugations in polar lobe surfaces, which are, in turn,
comparable to the thickness of the shell.  Hopefully, the model shape
in Figure 4 and Table 1 will be used to constrain future theoretical
investigations of the formation of bipolar nebulae in general, and the
Homunculus in particular.

\subsection{Some Details: Departures from Axial Symmetry}

Of course, the Homunculus lobes are not {\it perfectly} symmetric.
Departures from axial symmetry are most apparent in the near side of
the SE lobe, which appears somewhat scrunched or ``snub-nosed''
overall, and has a few large dents at some slit positions, especially
at the corners and right at the polar axis.  These features have been
mentioned before as they are apparent in images (e.g., Morse et al.\
1998).  This is only obvious in the SE lobe --- by contrast, the NW
lobe shape obeys axial symmetry to a higher degree, showing only a few
minor corrugations.

Overall, the H$_2$ shell appears smoother in the side walls than in
the polar caps.  The polar regions of the lobes show several clumps or
cells in the H$_2$ spectra, as well as in high-resolution {\it HST}
images (Morse et al.\ 1998).  These correspond to minor wiggles or
corrugations in the shape of the nebula -- but these deviations from a
smooth shape in the radial direction are smaller than sizes of cells
in images.  In other words, these are not three-dimensional ``clumps''
in the normal sense, but rather, fractal divisions of the outer
shell's surface.  They are more like irregular plates in the thin
shell, which bring to mind analogies with rapid cooling and thermal
instabilities or convection cells.  These analogies may be appropriate
during the optically thick, rapidly cooling phase after the material
was ejected.  The bright H$_2$ features, presumably tracing the
densest material, correspond to bright features in optical images,
rather than dark lanes.

The maximum radial expansion speed does not occur at the pole, but at
latitudes around 65--70\arcdeg\ (Fig.\ 5a).  This gives rise to a hint
of a trapezoidal shape to the lobes of the Homunculus.  This
trapezoidal shape is more obvious in the polar lobes of the LH; this
coincidence suggests a common physical mechanism at work.  One
possibility is that this results from extra ram pressure inside the
corners of the lobes, arising from the deflection of the post-eruption
wind that the inner side-walls of the polar lobes at an oblique angle.
Such effects are present in simulations of bipolar wind-blown bubbles
(e.g., Cunningham, Frank, \& Hartmann 2005; Frank et al.\ 1998).  This
extra ram pressure would not be related to the opening angle of the
shock cone in a binary system, since during the eruption it is likely
that the shock cone was collapsed onto the secondary star.

\section{THE GREAT ERUPTION AS A FUNCTION OF LATITUDE}

The Homunculus bears the imprint of the Great Eruption's mass loss as
a function of latitude.  Therefore, the shape and structure of the
H$_2$ emission in high-resolution spectra provide important insight to
the physics of the outburst, to the extent that H$_2$ is a good tracer
of the mass.

Figure 5 shows the latitude dependence of velocity, mass loss, and
mechanical energy in the Homunculus, based on H$_2$ emission.  The
velocity distribution follows directly from the shape of the nebula,
since it is a Hubble flow.  However, calculating these distributions
of mass and kinetic energy requires a few assumptions.  As a first
approximation, the thin outer H$_2$ shell of the Homunculus is taken
to be of uniform density with a constant width.  The apparent width in
Phoenix spectra is very nearly constant, with variations typically
less than a factor of 2.  Note that the thicker appearance of the
H$_2$ shell at slit offset positions NE2 and SW2 (see Fig.\ 1) is
somewhat misleading, since these slit positions slice through the
lobes at an oblique angle to their surfaces.  If this assumption
applies, then the mass in various parts of the shell is simply
proportional to its geometric surface area at each latitude.  While
the nebula is obviously not perfectly uniform, we will see below that
the important characteristics of the latitudinal distribution in the
nebula are not compromised by these minor effects.

To calculate various quantities as functions of latitude, the outer
shell was approximated as a series of adjacent loops or tori centered
on the polar axis of the nebula (see Figure 6).  The mass for each
representative latitudinal loop is given by 2$m_H n_H$ ($\pi
r^2$)(2$\pi$R$_H$), where R$_H$=R$_H$($\theta$) is the horizontal
radius of each torus measured from the polar axis (parallel to the
equator), and 2$r$ is the thickness of the torus, roughly equal to the
overall shell thickness of $\sim$600 AU (the extra factor of 2
accounts for the mass in both polar lobes).  Following the assumption
of constant shell thickness and uniform density, however, the fraction
of the total mass at each latitude
$f_M$($\theta$)=M($\theta$)/$\Sigma$M($\theta$) is simply proportional
to R$_H$.  This quantity is plotted in Figure 5b, and avoids the
uncertainty in the absolute physical density, shell thickness (as long
as it is constant), and total mass.  The fraction of the total kinetic
energy at each latitude
$f_{KE}$($\theta$)=M($\theta$)$v$($\theta$)$^2$/$\Sigma$[M($\theta$)$v$($\theta$)$^2$]
in Figure 5c is calculated for each torus from its value of
$f_M$($\theta$) and its radial expansion speed.  These rings were
regularly spaced along the surface of the nebula (Fig.\ 6), but not
spaced regularly in latitude.  Therefore, before plotting in Figure 5,
the values for $f_M$($\theta$) and $f_{KE}$($\theta$) were rebinned in
latitude intervals of 5\arcdeg.

Figure 5b indicates that most mass lost during the Great Eruption was
aimed to high latitudes.\footnote{A deficit of mass at low latitudes
raises interesting questions about the history of light escaping from
the Homunculus as the ejecta expanded and thinned in the decades after
the Great Eruption.  For example, Frew (2004) notes that the polar
lobes of the Homunculus were not seen until the 1940's (Gaviola 1950),
while early visual observers noted the appearance of ``nubeculei''
that are associated with equatorial features (Innes 1914, 1915; Voute
1925, van den Bos 1928; see also Smith \& Gehrz 1998).}  Only 25\% of
the total Homunculus mass was ejected at $0\arcdeg < \theta <
45\arcdeg$, with the remaining 75\% ejected between 45\arcdeg\ and the
pole.  The latitudinal distribution of the ejecta's kinetic energy is
even more extreme, with almost all the mechanical energy ($\sim$94\%)
escaping between 45\arcdeg\ and the pole.  The efficiency of imparting
mechanical energy to the ejecta seems to peak at latitudes around
60-65\arcdeg, while the peak mass loss occurs around 50--60\arcdeg.

For comparison, the latitudinal distribution of mass and kinetic
energy for a uniform spherical shell is shown by the dotted curves in
Figure 5, while Figure 7 shows the mass and kinetic energy per solid
angle (so that a spherical shell is a horizontal line).  The
differences between a spherical shell and the Homunculus are striking.
This latitudinal behavior strongly refutes the idea that a bipolar
nebula like the Homunculus was caused by an otherwise spherical
explosion or wind that is simply pinched by a pre-existing
circumstellar torus, because in that case we would expect to see more
mass at low latitudes (see also Dwarkadas \& Balick 1998 and Soker
2001).  Likewise, the mass loss during the Great Eruption could not
have been redirected toward the poles by deflection from its companion
star, because the amount of kinetic energy in the polar ejecta is
greater than the binding energy of the current putative binary system,
and more than three orders of magnitude greater than the available
energy in the companion's wind during the eruption.

Instead, the concentration of mass at high latitudes suggests that the
original explosion itself directed the mass and momentum toward the
poles.  An alternative way to express same result is that perhaps much
of the mass that would have been ejected at 0--40\arcdeg\ failed to
reach escape velocities and fell back onto the star.  In any case,
having the majority of the mass --- a significant fraction of the
total mass of the star -- lost at polar latitudes will effect the
star's angular momentum evolution.  A polar explosion will tend to
take away less than its share of angular momentum, leaving the
post-outburst star with higher angular momentum per unit mass.  This
is related to its presently-observed bipolar stellar wind (Smith et
al.\ 2003a).  Additionally, after ejecting its own massive LBV nebula,
there is evidence that AG Carinae is also left as a rapid rotator
(Groh et al.\ 2006).  Ultimately, without a circumstellar torus or
deflection by a companion, the bipolar shape must come from rapid
rotation of the primary star; in that view the role of a companion
star is relegated to helping to spin up the primary star through tidal
friction during repeated periastron passes (e.g., Smith et al.\
2003a).  If the mass loss is driven by radiation, then gravity
darkening on a rapidly rotating star will tend to drive most of the
mass flux toward high latitudes (e.g., Owocki et al.\ 1996, 1998;
Owocki \& Gayley 1997; Owocki 2003), providing a possible explanation
for the polar mass concentration.  However, it remains to be seen if
even a continuum-driven wind can fuel the required mass loss (Owocki
et al.\ 2004).

The latitudinal mass distribution has some interesting connections
with other famous bipolar nebulae, especially the circumstellar nebula
around SN1987a.  Both $\eta$ Car and SN1987a have nebulae with
equatorial and bipolar mass loss, but the two polar rings around
SN1987a are particularly vexing.  In the Homunculus of $\eta$ Car, we
see that most of the mass outflow was concentrated toward intermediate
latitudes around 65--70\arcdeg.  One can imagine that if the mechanism
that produced this mass concentration were pushed to some extreme
limit, it might produce a pair of rings as it flowed out from the
star.  While energy arguments rule out bipolar shaping by a companion
star in an eccentric 5.5 yr orbit that survives as the putative binary
system seen in $\eta$ Car today (see above), a more exotic model where
the 1843 Great Eruption involved the merger of a tight binary system
is admittedly more difficult to reject.  In fact, a recent theoretical
study of a merger scenario for the progenitor of SN1987a found that
most of the mass loss is directed toward mid latitudes, with a
significant deficit of mass loss near the equator (Morris \&
Podsiadlowski 2005).  On the other hand, if one wishes to explain
$\eta$ Car's Great Eruption as a merger of two massive stars, then it
is a mystery why the similar eruption of P Cygni in 1600 A.D.\ (see
Humphreys et al.\ 1999) ejected a spherical shell (Smith \& Hartigan
2006) and apparently did not leave that star as a rapid rotator.

Finally, an important caveat is needed.  The notorious equatorial
skirt of $\eta$ Carinae --- to be discussed in \S 7 --- is not
detected in near-IR H$_2$ emission, presumably because molecules
simply cannot survive or never formed there.  This equatorial spray of
ejecta seen in images is quite extensive and reaches large radii from
the star, comparable to some parts of the polar lobes.  Thus, one
should expect that the true velocity, mass, and energy distributions
in Figure 5 would be modified with a spike at $\theta$=0\arcdeg.  The
magnitude of this equatorial spike in the mass distribution is not
known.  Judging by the general absence of these equatorial features in
thermal-IR images (Smith et al.\ 2003b), however, the skirt likely
contains less than $\sim$0.5 M$_{\odot}$.

\section{MASS AND DENSITY IN THE H$_2$ SKIN}

A crude estimate of the expected mass density in the H$_2$ shell can
be made from the total mass of the Homunculus and the volume occupied
by the thin H$_2$ skin.  The H$_2$ skin corresponds to the cool 140 K
dust component that occupies the outer shell seen in thermal-IR
images, which Smith et al.\ (2003b) estimate to contain at least 11
M$_{\odot}$.  This 11 M$_{\odot}$ contains about 90\% of the total
mass in the Homunculus, but its actual value depends on the highly
uncertain gas:dust mass ratio, assumed to be 100.  As noted by Smith
et al.\ (2003b), this is most likely a lower limit to the mass, since
the ejecta of $\eta$ Car are depleted of important grain constituents
like C and O, and much of the Fe is evidently in the gas phase as
re-enforced by the spectra presented in this paper.  Therefore, the
density derived below will also be a lower limit.  The total volume of
the model H$_2$ skin in Figure 6 is V$_{H_2}$=4.6$\times$10$^{51}$
(r/300 AU)$^2$ cm$^{3}$.  Thus, the equivalent pure atomic hydrogen
density in the H$_2$ skin is given by

\begin{displaymath}
 n_H > \frac{ 11 M_{\odot} }{ m_H \  {\rm V}_{H_2} }  
 \simeq 3\times10^6 {\rm cm}^{-3}
\end{displaymath}

\noindent where $m_H$ is the proton mass and 2$r$=600 AU is the
assumed thickness of the H$_2$ skin.  Two points need to be stressed:
1) $n_H$$>$3$\times$10$^6$ cm$^{-3}$ is a lower limit because the
total mass estimate is a lower limit (Smith et al.\ 2003b) and because
parts of the shell are thinner than adopted here, and 2) this is an
{\it average} density, so densities above 10$^7$ cm$^{-3}$ may be
present in clumps even if the total mass estimate is accurate.

The assumption of constant average density throughout the H$_2$ skin
may be optimistic, of course, but given the very thin side walls of
the Homunculus lobes, the latitudinal mass trend in Figure 7 could
only be erased (flattened) if the density were of order 10$^8$
cm$^{-3}$ or more.  Corresponding column densities would be
$N_{H_2}\simeq$10$^{24}$ cm$^{-2}$, which would contradict the fact
that the side walls of the polar lobes are more optically thin than
the caps (e.g., Davidson et al.\ 2001; Allen \& Hillier 1993).  Such
high column densities would also contradict the much lower value of
$N_{H_2}\simeq$1.5$\times$10$^{23}$ cm$^{-2}$  from scattered hard
X-rays in the Homunculus (Corcoran et al.\ 2004).

Similar values of $n_H\simeq$10$^7$ cm$^{-3}$ have been derived by
Gull et al.\ (2005) from UV absorption along our sightline to the star
in the lowest ionization species at a radial velocity of --513 km
s$^{-1}$.  This rough agreement with the density value quoted here
gives independent confirmation of the very high mass estimate for the
Homunculus of more than 10 M$_{\odot}$ (Smith et al.\ 2003b).  In
fact, the volume of the H$_2$ shell derived from Phoenix spectra and
the higher density derived independently by Gull et al.\ would favor
an even higher mass for the Homunculus of perhaps $\sim$20
M$_{\odot}$.

\section{CLUES TO THE NATURE OF THE COMPLEX CORE AND THE BIZARRE
EQUATORIAL EJECTA}

While the shape, orientation, and density structure of the polar lobes
seem understandable in the framework of the double-shell structure
discussed here, the morphology of the messy equatorial skirt is still
perplexing.  However, the new Phoenix spectra from Gemini South shed
some light on this mystery.

One important clue is that the material in the extended equatorial
skirt shows no detectable infrared H$_2$ emission, despite residing at
large radii where it should be shielded by dense dusty structures in
the core region.  The features in the equatorial skirt, while
prominent in reflected light at visual wavelengths (Morse et al.\
1998; Duschl et al.\ 1995), have weak emission in the thermal-IR
(Smith et al.\ 2003b).  These facts suggest that the equatorial disk
has much lower density than the walls of the lobes, and so dust and
molecule formation was less efficient there in the decades after the
Great Eruption.  In fact, some equatorial features seem to be optical
illusions (Davidson et al.\ 2001).  Thus, the equatorial skirt
contains only a small fraction of the total mass in the Homunculus;
probably $\la$0.5 M$_{\odot}$, as noted above.

The spatial and kinematic structure in high-resolution spectra of
H$_2$ give the first clear picture of the shape and structure of the
Homunculus at low latitudes near equator where the two polar lobes
meet.  This region is hidden in images and optical spectra due to
obscuration from the foreground lobe, and the [Fe~{\sc ii}] emission
there is very confusing due to the Little Homunculus (Ishibashi et
al.\ 2003; Smith 2005) and structures near the star like the so-called
``Weigelt knots'' and other clumps (Weigelt et al.\ 1995; Davidson et
al.\ 1997; Smith et al.\ 2004; Smith 2005; Chesneau et al.\ 2005).
Thermal emission from dust traces an extremely complicated
distribution of interconnected clumps, arcs, and diffuse emission seen
most clearly in high resolution images at 2--8 $\micron$ (Chesneau et
al.\ 2005; Smith et al.\ 2002, 2003b; Smith \& Gehrz 2000; Rigaut \&
Gehring 1995).  This is the warmest dust (T$>$250 K) in the core of
the Homunculus, long thought to constitute some sort of a warm torus
(Hyland et al.\ 1979; Mitchell et al.\ 1983; Hackwell et al.\ 1986;
Smith et al.\ 1995, 1998; Polomski et al.\ 1999).  Two recent papers
with the best IR images to date suggested two very different
interpretations of these complex structures: 1) Smith et al.\ (2002)
proposed that they mark a distorted and disrupted torus or ring where
the two polar lobes of the Homunculus meet at the equator.  In this
view, the complex structures are a result of lower-density structures
getting swept back by the post eruption stellar wind, and higher
density clumps resisting that expansion.  2) Chesneau et al.\ (2005)
proposed the existence of a new structure called the ``Butterfly
Nebula'', which was a dusty bipolar nebula different from the Little
Homunculus, and which had the dust reaching back toward the star along
the polar axis.  The new H$_2$ spectra distinguish clearly between
these two hypotheses.

Figure 8 demonstrates the correspondence between the structures seen
in a high-resolution map of the warm dust column density near the star
from Smith et al.\ (2003b) and the spatial and kinematic structure of
H$_2$ emission in spectra.  Where the thin H$_2$ walls of the
Homunculus meet at the equator, they do not reach all the way in to
the star, but instead leave a gap in the apparent distribution of
H$_2$ along the slit.  This gap is slightly different at each slit
position, and is noted in each spectral image.  Each of these gaps is
superposed on top of the IR continuum image, where we see that indeed,
the spatial limits of H$_2$ emission correspond nicely to the empty
interior of the so-called ``Butterfly Nebula'' or torus.  The
agreement between H$_2$ in the walls of the Homunculus and the dust
structures prove that this really is an equatorial torus or ring where
the polar lobes meet at the equator, rather than the separate nebular
structure proposed by Chesneau et al.  The toroidal interpretation
also gives a more natural explanation for the deficit of IR emission
within its boundaries (Chesneau et al.\ 2005).  No kinematic evidence
for the putative butterfly structure is seen in spectra, and the dust
structures are not coincident with the Little Homunculus either (Fig.\
8).  The question then is: if these structures represent parts of an
equatorial torus or ring, why does it exhibit such prominent azimuthal
asymmetry (clumps, arcs, etc.), while the rest of the H$_2$-emitting
portions of the Homunculus, by comparison, show such well-ordered
axisymmetric structure?

The strange clumps and arcs in this equatorial dust structure -- and
the lack of similar features in the polar lobes -- can be understood
as follows (Smith et al.\ 2002): In the 160 yr since the Homunculus
was ejected in the Great Eruption, it has been followed by the
post-eruption stellar wind.  At the poles, the speed for the bulk of
the post eruption wind is about 600 km s$^{-1}$ (traced by the deepest
part of the H$\alpha$ P Cygni absorption; Smith et al.\ 2003a), which
is almost identical to the expansion speed of the Homunculus (Fig 5a).
With such a small $\Delta v$, we expect little or no hydrodynamic
influence from the wind.  The equator presents a very different story.
Although the present-day stellar wind is bipolar, it appears to
maintain relatively high wind speeds of $\sim$400 km s$^{-1}$ at the
equator (Smith et al.\ 2003a), whereas the pinched waist of the
Homunculus seen in H$_2$ has very slow velocities of only a few dozen
km s$^{-1}$.  This large value of $\Delta v$ right at the equator
should lead to Rayleigh-Taylor or nonlinear thin-shell instabilities
(Vishniac 1994) in the torus where the polar lobes meet, much like any
other dense ejecta shell followed by a faster wind (e.g.,
Garcia-Segura et al.\ 1996).\footnote{The much lighter and slower
shell around P Cygni is a good example of this (Smith \& Hartigan
2006).}  The Homunculus is inherently clumpy at all latitudes, perhaps
even more so at the equator (ud Doula \& Owocki 2002). Since the ram
pressure of the incident wind depends on ($\Delta v$)$^2$, the
post-eruption wind is most influential at the equator.  Any clumps in
this ring will resist the ram pressure of the wind more than the less
dense regions between them.  The inter-clump sections of the torus
would be swept back, leading to structures akin to dust pillars in
H~{\sc ii} regions.  In other words, the values in the first row of
Table 1 are azimuthally dependent.  This is essentially a more extreme
version of the Rayleigh-Taylor instabilities that are thought to have
led to the early development of hot spots in the famous equatorial
ring around SN1987a (e.g., Michael et al.\ 1998).

In the inter-clump regions of the torus, where the wind pushes
outward, there may be an interesting link to the unusual structures
seen in $\eta$ Car's equatorial skirt.  (The equatorial skirt is the
more extended disk with peculiar radial streamers or spokes that seem
to point back to the star; Duschl et al.\ 1995; Morse et al.\ 1998.)
In the inter-clump regions of the torus that are pushed to larger
radii, light may more easily leak through to illuminate outer regions
of the skirt (i.e. searchlights like the radial features in the Egg
Nebula; Sahai et al.\ 1998).  In some extreme cases, the stellar wind
or ejecta from the Great Eruption or the 1890 eruption may even have
broken through the Homunculus.  Some material in the skirt is indeed
seen to have kinematics consistent with ejection during the 1890
eruption (Davidson et al.\ 2001; Smith 2002a), and some more extended
equatorial features like the NN Jet and the S Condensation appear to
have originated in the Great Eruption (Morse et al.\ 2001).  These
searchlights or ejecta breaking through the Homunculus in the equator
provide a natural explanation for the quasi-radial spokes or streamers
seen in the equatorial skirt.  These features in the skirt may be
largely an illumination effect, while the true distribution of matter
in the equatorial disk may be considerably more azimuthally symmetric
than it appears.

\section{CONCLUSIONS}

New high-resolution, long-slit spectra of H$_2$ $\lambda$21218 and
[Fe~{\sc ii}] $\lambda$16435 from the Phoenix spectrograph on Gemini
South provide new clues to the detailed structure of the Homunculus.
The main conclusions of this study are summarized as follows:

1.  The relative spatio-kinematic structure of infrared H$_2$ and
    [Fe~{\sc ii}] emission gives an independent confirmation of the
    double-shell structure for the polar lobes of the Homunculus
    proposed on the basis of thermal-IR images (Smith et al.\ 2003b).
    The inner [Fe~{\sc ii}] layer corresponds to the warmer (200 K)
    inner dust layer that traces only about 10\% of the mass of the
    Homunculus, while the outer H$_2$ skin traces the coolest dust
    component (140 K) that contains the remaining 90\% of the nebular
    mass.

2.  The high-resolution spectra of H$_2$ show for the first time how
    thin and surprisingly smooth the outer molecular skin of the
    Homunculus is.  If the H$_2$ skin's $\Delta$R/R traces $\Delta$t/t
    of the mass ejection, then the most important mass-loss phase
    during the eruption is constrained to be about 5 yr or less.  The
    corresponding mass-loss rate is so high that the mechanical
    luminosity would have greatly exceeded the radiative luminosity at
    that time, implying that the Great Eruption was more of an
    explosion than a wind.

3.  A lower limit to the density within the H$_2$ layer, based on a
    lower limit to the mass of the Homunculus (Smith et al.\ 2003b)
    and the apparent volume of the H$_2$ skin, is $n_H \ga 10^{6.5}$
    cm$^{-3}$.

4.  Since H$_2$ emission traces most of the mass in the Homunculus as
    well as the dominant dust scattering layer seen as a reflection
    nebula in images, H$_2$ spectra give the first definitive picture
    of the 3D shape of the Homunculus, which is given in Table 1.
    Earlier estimates of the shape and orientation traced the inner
    [Fe~{\sc ii}]-emitting layer, which has a different size and
    structure.

5.  The model shape combined with the known age of the nebula gives a
    distance of 2350$\pm$50 pc to $\eta$ Car, and an inclination of
    $i$=41\arcdeg$\pm$0$\fdg$5 for the polar axis.

6.  Deviations from a smooth symmetric shape are minor -- generally
    comparable to the thickness of the lobes.  The redshifted NW lobe
    appears more symmetric than the approaching SE lobe, which shows
    more prominent dents and corrugations.

7.  The H$_2$ distribution yields our first estimate of the
    latitudinal dependence of the velocity, mass loss, and kinetic
    energy of the Great Eruption.  About 75\% of the mass and more
    than 90\% of the mechanical energy was released at latitudes
    between 45\arcdeg\ and the pole.

8.  This latitudinal mass distribution rules-out a model where the
    bipolar shape of the nebula arises from an otherwise spherical
    explosion being constricted by a pre-existing circumstellar torus.
    The kinetic energy of the polar ejecta rule out a model where the
    ejecta were deflected toward the poles by a companion star,
    because the putative companion could not have supplied the
    required energy.  This means that the ejecta were directed toward
    the poles by the explosion itself.  A merger event is not as easy
    to rule out.

9.  The spatio-kinematic structure of H$_2$ emission at the pinched
    waist of the nebula helps explain the unusual and complex
    structures seen in high-resolution IR images as an equatorial
    torus disrupted by the post-eruption wind.  The radial structures
    in the equatorial skirt seen in visual-wavelength images are the
    result of either beams of light or ejecta that have broken through
    the inner dust torus.

\acknowledgments \scriptsize

I thank Verne Smith and Sybil Adams for assistance during the Gemini
South observing run, and Keivan Stassun for diligently obtaining
numerous dark frames with Phoenix earlier in the same observing
nights.  I also thank Kris Davidson and Stan Owocki for comments on
the manuscript and numerous discussions.  I was supported by NASA
through grant HF-01166.01A from the Space Telescope Science Institute,
which is operated by the Association of Universities for Research in
Astronomy, Inc., under NASA contract NAS5-26555.


\clearpage
\begin{deluxetable}{lcccc}
\tabletypesize{\scriptsize}
\tighten\tablenum{1}\tablewidth{0pt}
\tablecaption{Homunculus Model Shape}
\tablehead{
  \colhead{Latitude} &\colhead{Radius} &\colhead{Exp.\ Vel.}
  &\colhead{X size}  &\colhead{Y size} \\
  \colhead{(degrees)} &\colhead{(AU)} &\colhead{(km s$^{-1}$)}
  &\colhead{(AU)}  &\colhead{(AU)}
}
\startdata

      0.0	      &2100	      &62	      &2100           &0.0 \\
      2.0	      &2349	      &70	      &2347           &82  \\
      4.0	      &2997	      &89	      &2990	      &209 \\
      8.6	      &3709	      &111	      &3668	      &555 \\
      13.9	      &4365	      &130	      &4237	      &1049 \\
      18.7	      &4961	      &148	      &4698	      &1594 \\
      21.9	      &5608	      &167	      &5202	      &2097 \\
      25.3	      &6276	      &187	      &5675	      &2681 \\
      28.6	      &6858	      &205	      &6021	      &3284 \\
      31.6	      &7422	      &222	      &6323	      &3887 \\
      34.2	      &7999	      &239	      &6613	      &4501 \\
      36.5	      &8654	      &258	      &6957	      &5149 \\
      38.7	      &9259	      &276	      &7225	      &5790 \\
      40.7	      &9935	      &297	      &7530	      &6481 \\
      42.1	      &10545	      &315	      &7824	      &7071 \\
      43.9	      &11200	      &334	      &8075	      &7761 \\
      45.5	      &11880	      &355	      &8327	      &8474 \\
      46.8	      &12487	      &373	      &8551	      &9100 \\
      48.3	      &13156	      &393	      &8756	      &9820 \\
      49.4	      &13823	      &413	      &8990	      &10500 \\
      50.5	      &14472	      &432	      &9210	      &11164 \\
      51.6	      &15044	      &449	      &9336	      &11798 \\
      52.6	      &15622	      &466	      &9481	      &12417 \\
      53.8	      &16275	      &486	      &9624	      &13126 \\
      54.7	      &16886	      &504	      &9753	      &13786 \\
      55.8	      &17503	      &523	      &9837	      &14478 \\
      56.9	      &18097	      &540	      &9869	      &15169 \\
      58.1	      &18716	      &559	      &9883	      &15895 \\
      59.1	      &19255	      &575	      &9878	      &16528 \\
      60.2	      &19779	      &591	      &9819	      &17170 \\
      61.4	      &20374	      &608	      &9739	      &17896 \\
      62.9	      &20898	      &624	      &9526	      &18601 \\
      64.3	      &21323	      &637	      &9249	      &19213 \\
      65.9	      &21641	      &646	      &8850	      &19749 \\
      67.7	      &21906	      &654	      &8327	      &20262 \\
      69.5	      &22014	      &657	      &7710	      &20621 \\
      71.3	      &22012	      &657	      &7060	      &20850 \\
      73.1	      &22001	      &657	      &6400	      &21050 \\
      74.9	      &21967	      &656	      &5740	      &21204 \\
      76.6	      &21932	      &655	      &5100	      &21332 \\
      78.3	      &21892	      &654	      &4450	      &21435 \\
      79.9	      &21873	      &653	      &3800	      &21540 \\
      81.8	      &21833	      &652	      &3100	      &21613 \\
      83.8	      &21774	      &650	      &2370	      &21645 \\
      85.6	      &21730	      &649	      &1660	      &21667 \\
      87.3	      &21711	      &648	      &1005	      &21688 \\
      89.1	      &21690	      &648	      &335	      &21688 \\

\enddata
\tablecomments{See Figure 6.  The last two columns are the Cartesian
  sizes of each point in the model shape in AU.}
\end{deluxetable}

\clearpage
\begin{figure}
\epsscale{0.5}
\plotone{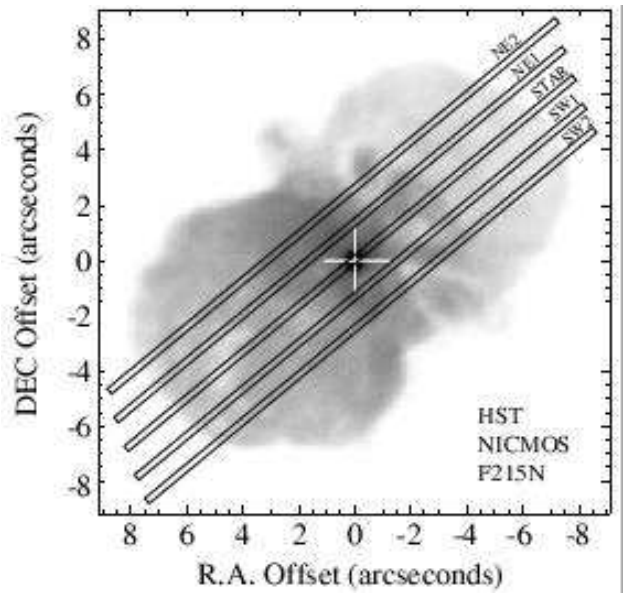} 
\figcaption{Phoenix slit aperture positions superposed on a
2~$\micron$ HST/NICMOS image of $\eta$ Car from Smith \& Gehrz
(2000).}
\end{figure}

\begin{figure}
\epsscale{0.8}
\plottwo{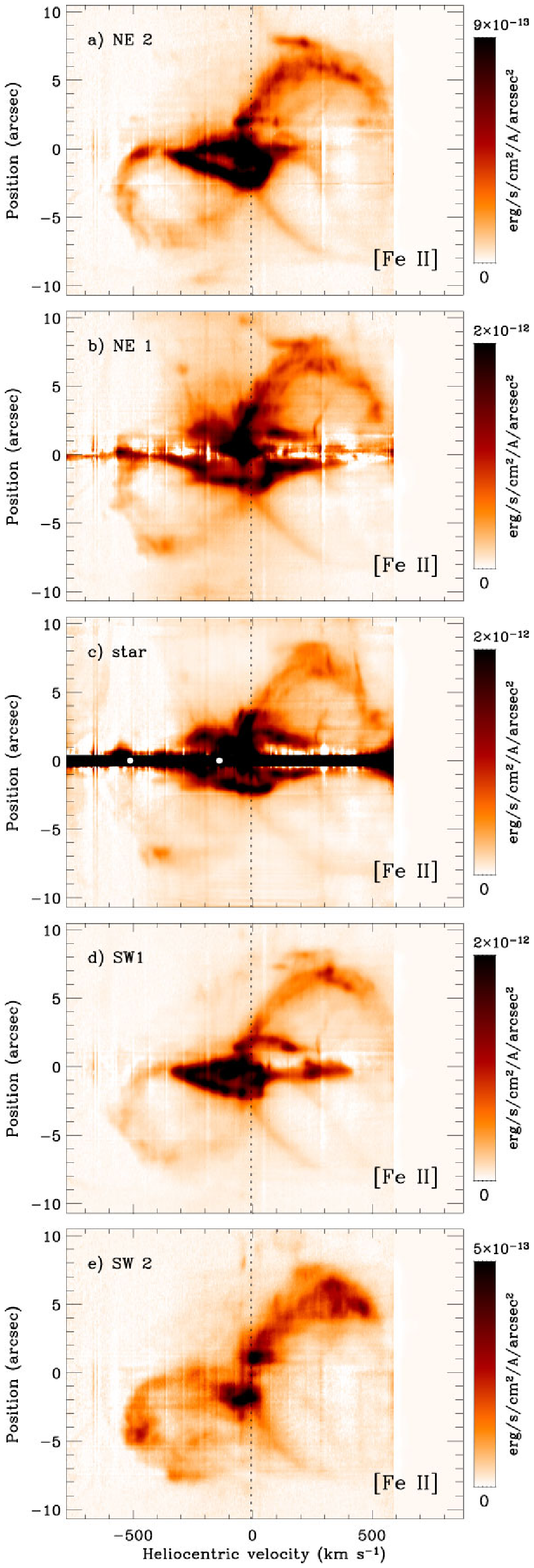}{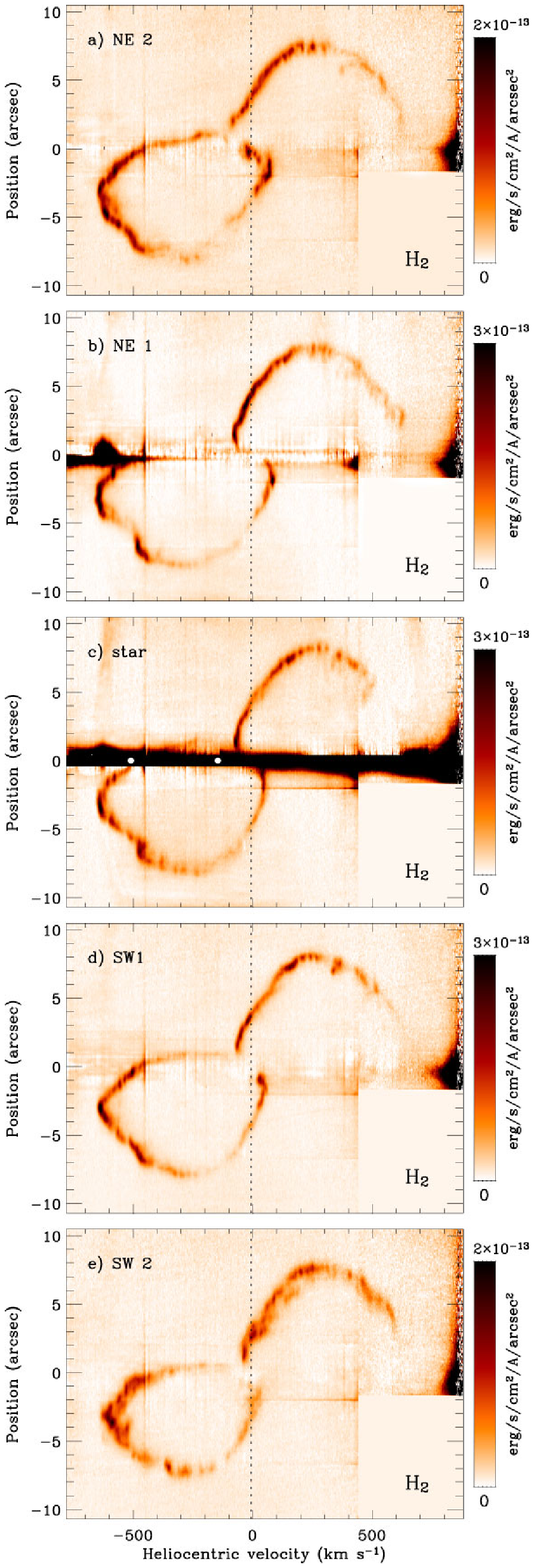}
\caption{Kinematic structure of [Fe~{\sc ii}] $\lambda$16435 emission
  (LEFT) and H$_2$ $v$=1--0 S(1) $\lambda$21218 emission (RIGHT),
  corresponding to the five different slit positions in Figure 1.
  Reflected continuum emission has been subtracted with a linear fit,
  but some residuals remain, especially near the bright central star.
  The dashed line shows the systemic velocity of --8.1 km s$^{-1}$
  (heliocentric).  Small white dots in slits passing through the
  central star mark the --513 and --146 km s$^{-1}$ absorption
  components seen in UV spectra.}
\end{figure}

\begin{figure}
\epsscale{0.72}
\plotone{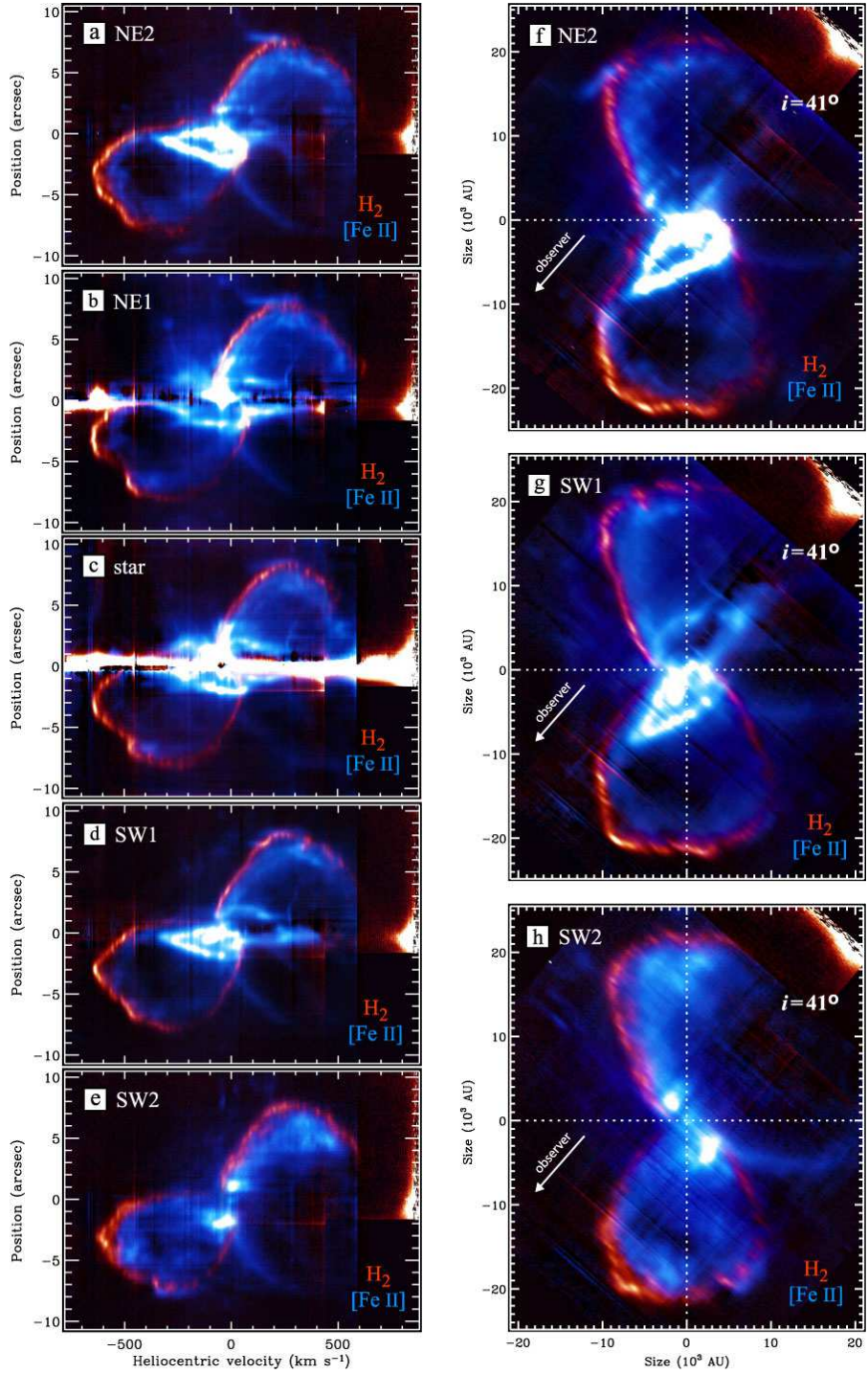}
\caption{(a--e) Same as for Figure 2, but showing both [Fe~{\sc ii}]
  $\lambda$16435 (blue) and H$_2$ $v$=1--0 S(1) $\lambda$21218 (red)
  simultaneously.  Panels f, g, and h show the data from panels a, d,
  and e, respectively, but with the velocity and spatial scales
  converted to AU (assuming an age of 160 yr and a distance of 2350
  pc), and rotated to correct for an inclination of
  $i$=41$\fdg$0$\pm$0$\fdg$5.}
\end{figure}

\begin{figure}
\epsscale{0.5}
\plotone{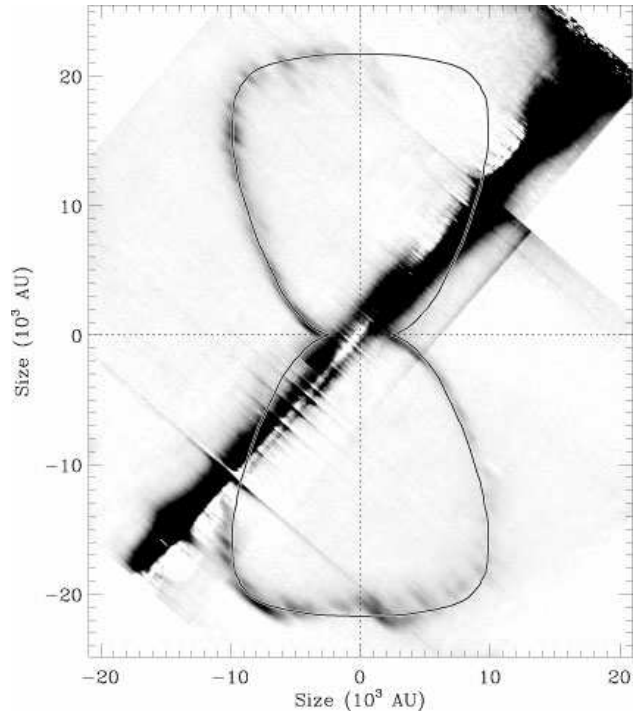}
\caption{The model shape from Table 1 is plotted over H$_2$
  $\lambda$21218 emission from the Homunculus.  The grayscale H$_2$
  emission is based on the same data as in Figure 3$c$, but the
  velocity scale was converted to AU by the factor appropriate for an
  age of 160 yr and a distance of 2.35 kpc (as in Figs 3f, g, and h).
  This yielded an ``image'' of a slice through the polar lobes and the
  star along our line of sight. The resulting image was then rotated
  by the apparent inclination angle of the system of 41\arcdeg \ so
  that the polar axis appears vertical here.  The unpleasant dark
  feature running diagonally through the image is residual continuum
  emission from the subtraction of the bright central star, which was
  saturated.}
\end{figure}

\begin{figure}
\epsscale{0.45}
\plotone{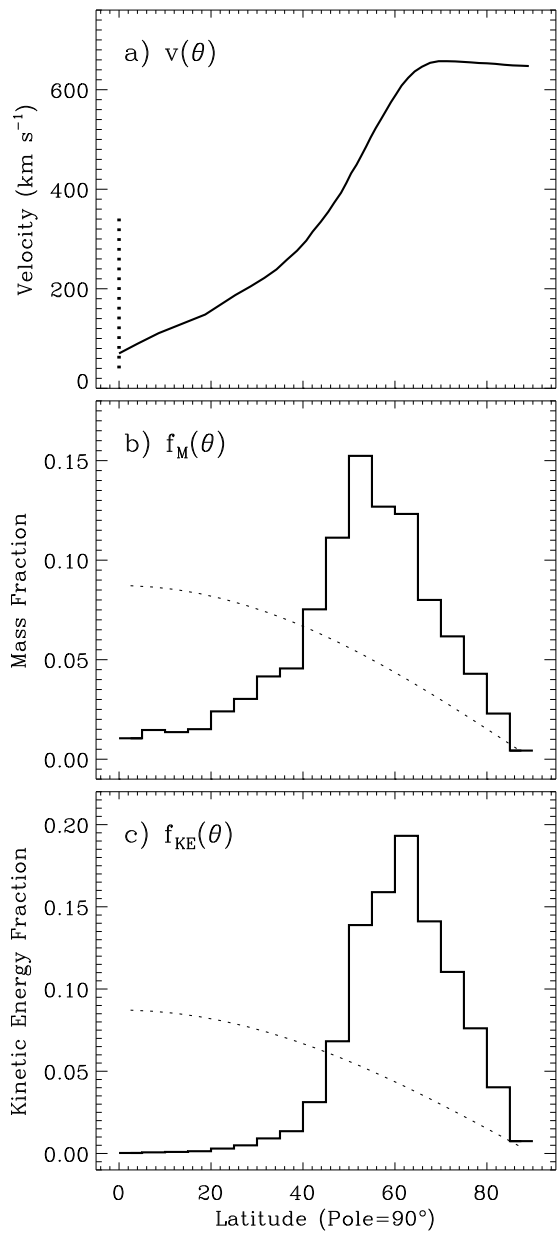}
\caption{Physical properties of the Great Eruption as a function of
  latitude, as traced by H$_2$ emission from the polar lobes.  The
  ejection velocity (a) is calculated from the shape in Figure 4 and
  Table 1, with an age of 160 yr.  The dashed vertical line at
  $\theta$=0$\arcdeg$ just accounts for the presence of the equatorial
  skirt inferred from images, since it does not appear in H$_2$
  emission.  (b) Fraction of the total mass ($f_M$) as a function of
  latitude in 5\arcdeg\ bins, assuming constant shell density and
  thickness across the surface area of the Homunculus.  (c) The
  fraction of the total kinetic energy in each 5\arcdeg\ latitude bin,
  from the corresponding mass fraction and velocity above.  The dotted
  curves in Panels b and c correspond to the same quantities for a
  uniform spherical shell.}
\end{figure}

\begin{figure}
\epsscale{0.5}
\plotone{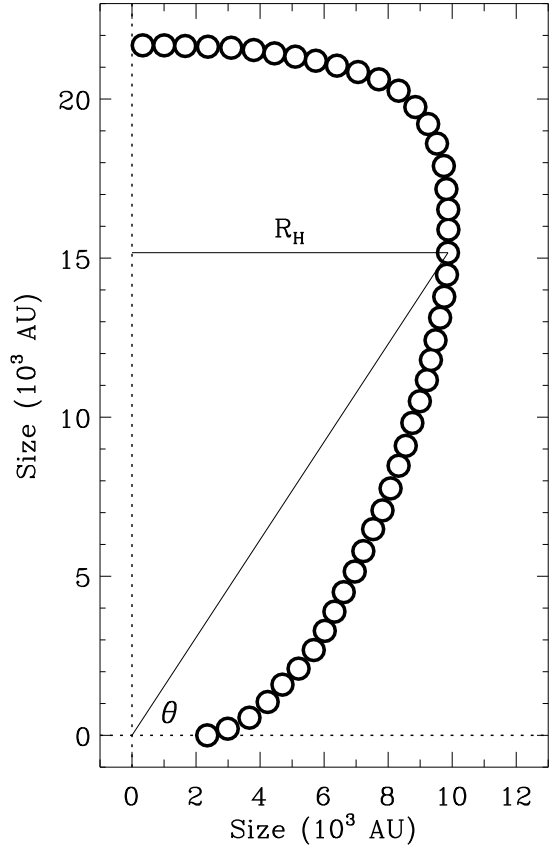}
\caption{Subdivision of a uniform thin shell into a series of
  regularly-spaced tori matching the model shape of the Homunculus,
  each with radius R$_H$ from the polar axis.  Each torus has the same
  cross-sectional area, but each takes up a different fraction of the
  surface area of the Homuculus, proportional to R$_H$.  This model
  was used to calculate the latitude dependence of mass and kinetic
  energy in Figure 5.  This sort of model accounts in a crude way for
  an effective filling factor due to clumps in the H$_2$ skin.}
\end{figure}

\begin{figure}
\epsscale{0.45}
\plotone{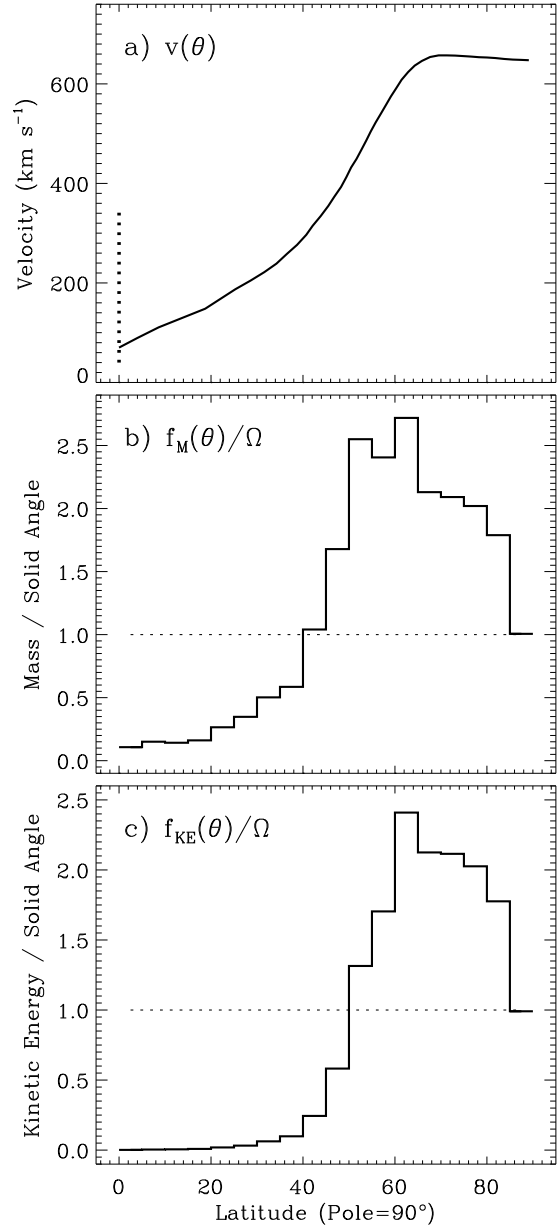}
\caption{Same as Figure 5, but showing the relative mass and kinetic
  energy fractions per solid angle, normalized to the value at the
  pole.  In these plots, a spherically-symmetric shell has the same
  mass and kinetic energy per solid angle at all latitudes.}
\end{figure}

\begin{figure}
\epsscale{0.95}
\plotone{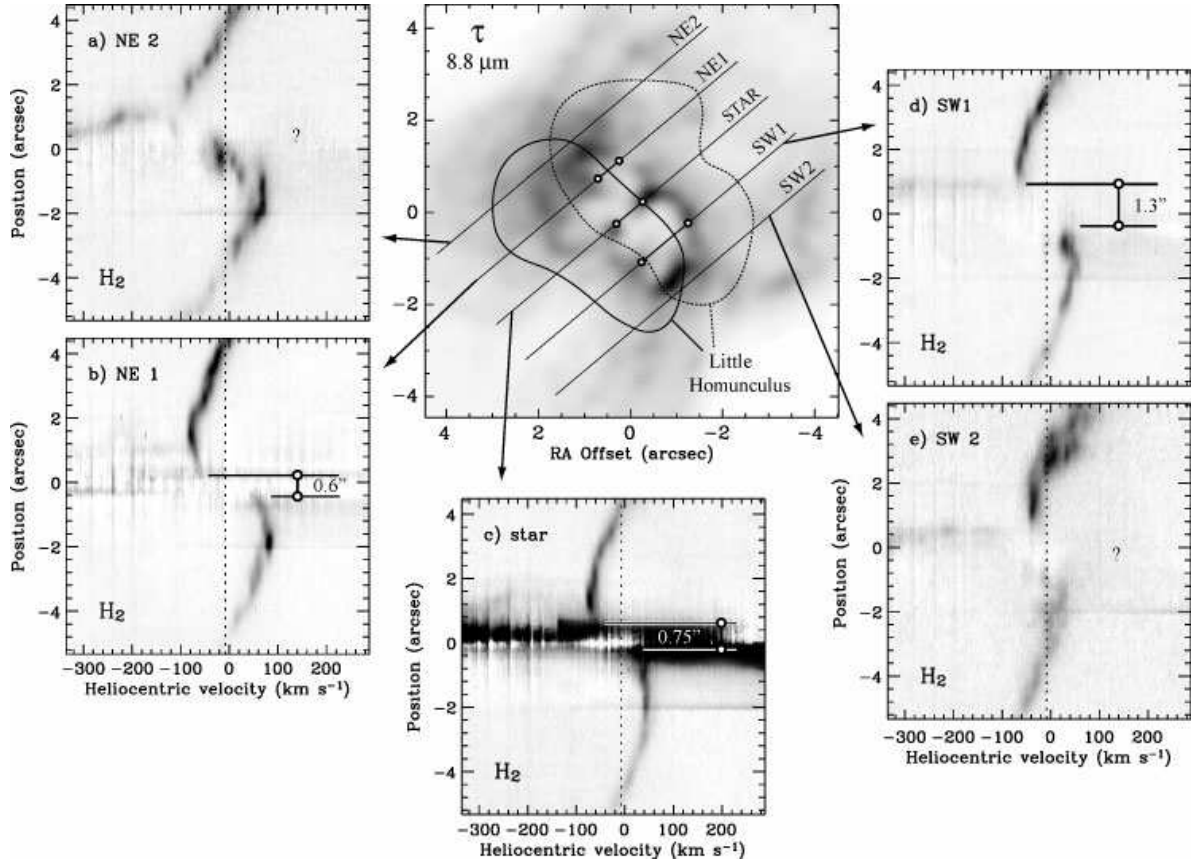}
\caption{The image in the central panel is the warm dust column
  density map (8.8 $\micron$ optical depth from Smith et al.\ 2003b),
  showing the bright dust structures near the core that have been
  interpreted as either a dust torus or a ``Butterfly Nebula''.
  Panels a--e are the same as the H$_2$ spectra in Figure 2a--e, but
  zoomed-in on the central structures.  The purpose of this figure is
  to illustrate the correspondence between spatial gaps in H$_2$
  emission in spectra and the size of the cavity in the interior of
  the dust torus.  The NE2 and SW2 slit positions do not cut through
  the interior of the torus, but for NE1, the star, and SW1, the gap
  in H$_2$ emission in each panel is superposed on the dust image as
  dots connected by a short solid line.  The gaps in H$_2$ emission
  match the interior of the dust torus, proving that this is an
  equatorial structure where the two polar lobes meet, rather than a
  bipolar ``Butterfly Nebula''.  The spatial extent of the Little
  Homunculus from Smith (2005) is shown as well, indicating that the
  dust structures do not arise in the Little Homunculus.}
\end{figure}

\end{document}